\documentclass[conference]{IEEEtran}
\IEEEoverridecommandlockouts
\usepackage{cite}
\usepackage{amsmath,amssymb,amsfonts}
\usepackage{algorithmic}
\usepackage{graphicx}
\usepackage{textcomp}
\usepackage{setspace}
\usepackage{xcolor}
\usepackage[flushleft]{threeparttable}
\def\BibTeX{{\rm B\kern-.05em{\sc i\kern-.025em b}\kern-.08em
    T\kern-.1667em\lower.7ex\hbox{E}\kern-.125emX}}

\newcommand{\asol}[1]{{\color{blue}{#1}}}
\newcommand{\lliu}[1]{{\color{brown}{#1}}}

\newcommand{\paren}[1]{\left( #1 \right)}
\newcommand{\cbrace}[1]{\left\{#1\right\}}

\newcommand{\real}{\mathbb{R}}

\newcommand{\vx}{\vec{x}}
\newcommand{\vX}{\vec{X}}

\newcommand{\squeezeup}[1]{\vspace{-#1pt}}

\usepackage{comment}
\usepackage{subfigure}
\usepackage{centernot}
\usepackage{etoolbox}
\makeatletter
\patchcmd{\@makecaption}
  {\scshape}
  {}
  {}
  {}
\makeatother

\begin{document}

\title{Neural Network Coding\\
}

\author{{\bf  Litian Liu, Amit Solomon, Salman Salamatian, Muriel M\'edard }\\
Research Lab of Electronics, MIT, Cambridge, MA \\
Email: \{litianl, amitsol, salmansa, medard\}@mit.edu\squeezeup{10}
}

\maketitle

\begin{abstract}
In this paper we introduce Neural Network Coding (NNC), a data-driven approach to joint source and network coding.
In NNC, the encoders at each source and intermediate node, as well as the decoder at each destination node, are neural networks which are all trained jointly for the task of communicating correlated sources through a network of noisy point-to-point links.
The NNC scheme is application-specific and makes use of a training set of data, instead of making assumptions on the source statistics.
In addition, it can adapt to any arbitrary network topology and power constraint.
We show empirically that, for the task of transmitting MNIST images over a network, the NNC scheme shows improvement over baseline schemes, especially in the low-SNR regime.
\end{abstract}


\vspace{-1em}
\section{Introduction}\label{sec:intro}

In recent years, there has been an increased effort towards data-driven code constructions, e.g.,~\cite{o2016learning,o2017introduction,felix2018ofdm,bourtsoulatze2019deep,farsad2018deep}.
As opposed to the traditional methods, data-driven approaches make no assumptions on the source statistics,
which leads to a significant improvement when looking at complex sources of data such as images.
Instead, such methods aim at discovering efficient codes by making use of a (potentially large) pool of data, in conjunction with a learning algorithm.
A successful candidate for the latter is Neural Network (NN), which has tremendous potential in many application domains.
Many of the previous works have focused on the point-to-point communication problem, under various channels. For example, learning the physical layer representation is studied for single-input and single-output (SISO) system in~\cite{o2016learning}, multiple input and multiple output (MIMO) system in~\cite{o2017introduction} and orthogonal frequency-division multiplexing (OFDM) system in~\cite{felix2018ofdm}. NN-based joint source channel coding (JSCC) is proposed for images in \cite{bourtsoulatze2019deep}, and for text in \cite{farsad2018deep}.
Though traditional techniques are optimal in the asymptotic regimes, in practical scenarios, it was shown in~\cite{o2016learning,o2017introduction,felix2018ofdm,bourtsoulatze2019deep,farsad2018deep} that NN-based methods were competitive, and even out-performed state of the art methods in some signal-to-noise ratio (SNR) regimes.

\begin{figure}
    \centering
    \includegraphics[width=1 \columnwidth]{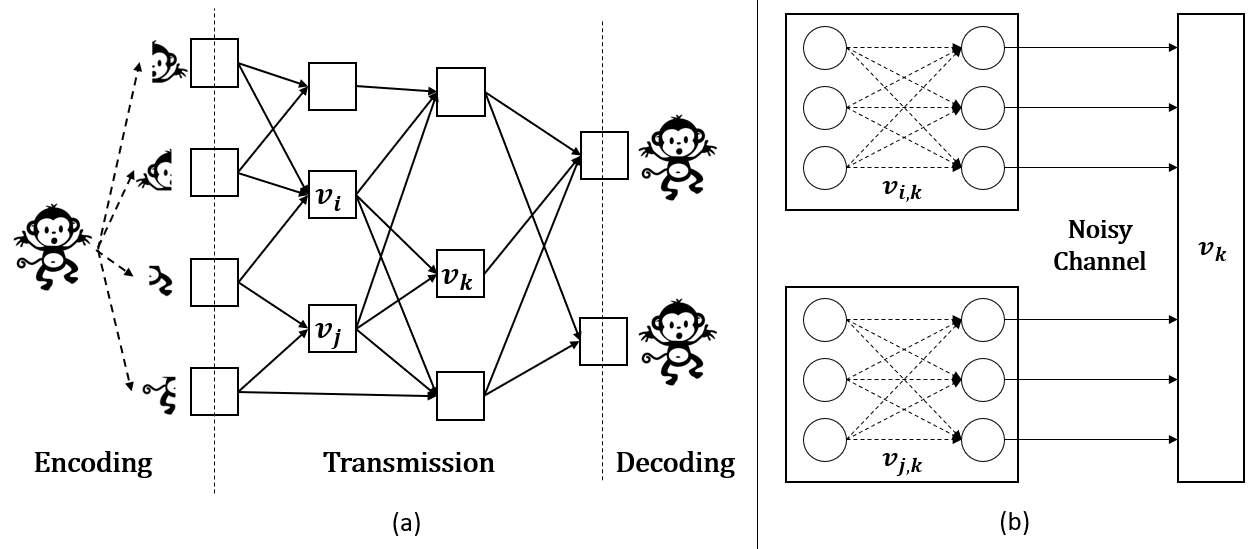}
    \caption{NNC overview. (a) Overview with $N = 4$ source nodes and $M = 2$ destination nodes. (b) Nodes $v_i$, $v_j$ to node $v_k$.\squeezeup{10}}
    \label{fig:general}
\end{figure}

In this work, we shift the scope to the network perspective, and make a first step in broadening our understanding of the benefits and challenges of data-driven approaches for a networked communication scheme. More precisely, we look at the problem of multi-casting correlated sources over a network of point-to-point noisy channels. The problem of multi-casting correlated sources has a long history starting from the seminal work of Slepian and Wolf (S-W) \cite{slepian1973noiseless}. 
In the S-W problem, a direct edge exists from each source to the destination.
Csisz\'{a}r showed in \cite{csiszar1982linear} that linear codes are sufficient to achieve the S-W bound.
For the scenario where correlated sources have to be communicated over a point-to-point network, the achievable rates were derived in \cite{song2001network}. It has been shown in \cite{ho2006random} that random linear network coding can be used for multi-casting correlated sources over a network, with error exponent generalizing the linear S-W coding \cite{csiszar1982linear}. 
However, for arbitrarily correlated sources, either maximum a posterior (MAP) decoders or minimum entropy (ME) decoders are required for random linear network coding.
Despite the efforts toward reducing the complexity of ME or MAP decoders\cite{maierbacher2009practical}\cite{coleman2005towards}, it is in general NP-hard to find a ME or MAP solution. 
The decoding complexity of the proposed joint coding scheme of \cite{ho2006random} has motivated Ramamoorthy et al. \cite{ramamoorthy2006separating} to seek a separation between source and network coding.
Such separation would allow the use of efficient source codes at the sources, followed by efficient network codes, thus allowing an overall efficient code construction.
Unfortunately, such separation turned out to be impossible in general. 
Following this, joint source and network coding schemes with low-complexity decoders have been proposed, but those suffer from restricted settings.
For example, in \cite{wu2009practical} sources are restricted to be binary and in \cite{lee2007minimum} the number of sources is restricted to be two. 
To the best of our knowledge, there is no existing practical joint source and network coding scheme for multi-casting arbitrarily correlated real-life signals over a point-to-point network of arbitrary topology.

To this end, we propose an \emph{application-specific} novel code construction based on NN, which we refer to as \emph{Neural Network Coding}, or NNC for short.
The NNC scheme has the following main benefits: (a) it makes no assumptions on the statistics of the source, but rather makes use of a seed dataset of examples; (b) it is an end-to-end communication scheme, or equivalently, a joint source and network coding scheme; (c) it has practical decoders; (d) it can be applied to any network topology; and (e) it can also be used with various power constraints at each of the source and intermediate nodes. 
Figure~\ref{fig:general}(a) demonstrates a NNC applicable scenario, where arbitrarily correlated signals are multi-cast over a communication network. In the network, there are four source nodes and two destination nodes. In fact, NNC can find wide applications from Internet of Things (IoT) to autonomous driving, where correlated signals generated by distributed measurements need to be combined for processing at central processors. In IoT and autonomous driving, signals are task-specific, leading to an efficient \textit{application-specific} scheme. Also, latency, bandwidth and energy can be extremely constrained in both cases, precluding computationally demanding long-block length source and channel coding techniques, let alone joint multi-casting scheme with ME or MAP decoders~\cite{ho2006random}. 

In NNC, the encoders at the source nodes and intermediate nodes, as well as the decoders at the destination nodes, are NNs as shown in Figure~\ref{fig:general}(b). 
The resulting network code construction is jointly designed with the encoding phase and decoding phase of the transmission, where real-valued input signals are mapped into channel inputs, and channel outputs are reconstructed into real-valued signals.
The end-to-end NNC scheme can be optimized through training and testing offline over a large data set, and can be readily implemented.
Of particular interest for these codes, is the \emph{power-distortion trade-off} they achieve.
In other words, for a given topology and power constraints on the nodes, what is the expected distortion that the code achieves, where the distortion measure is specified.
NNC is reminiscent of the auto-encoder\cite{goodfellow2016deep} structure. 
An auto-encoder is a NN trained to minimize the distortion, e.g. Mean Square Error (MSE), between its output and input.
The end-to-end NN structure that results from NNC scheme is similar to the auto-encoder mentioned above, with some additional constraints imposed by the topological structure of the physical communication network.
Our experimental results showed the benefit of having non-linear code construction in this setup.
Furthermore, we illustrate through experiments on images that NNC achieves better performance compared to a separated scheme based on a compression scheme (JPEG \cite{wallace1992jpeg}), followed by capacity achieving network coding. While still in its infancy, we believe that NNC and its variants may pave the way to an efficient way to exploit non-linearity in codes, which appears to be an important component to more complex networked settings.

The rest of the paper is organized as follows. Section~\ref{sec:model} describes our system model and Section~\ref{sec:method} presents our design principle. Section~\ref{sec:experiment} studies the power-distortion trade-off of NNC under a variety of network conditions. Section~\ref{sec:conclusion} summarizes the paper and discusses possible extensions.

\section{System model}\label{sec:model}

Throughout the paper, we use $x,\vx, X, \vX$ to denote a scalar, a vector, a random variable, and a random vector respectively. 
We model the communication network as an acyclic directed graph $G = (\mathcal{V}, \mathcal{E})$ \cite{cormen2009introduction}. Elements of $\mathcal{V}$ are called nodes and elements $(i,j)$ of $\mathcal{E}$ are called links. Each of the links $(i,j)$ is assigned an energy constraint $p_{i,j} \geq 0$, which specifies the maximum signal energy that can be sent from node $v_i$ to node $v_j$. We consider two disjoint subsets $\mathcal{S}$, $\mathcal{D}$ of $\mathcal{V}$, where each element in $\mathcal{S}$ is called a source node and each element in $\mathcal{D}$ is called a destination node. Let $N$ and $M$ denote $|\mathcal{S}|$ and $|\mathcal{D}|$ respectively.
We consider $n$ virtual sources $ \{ s_i \}_{i=1}^n$ located at $N$ source nodes. Each $s_i$ generates a random variable $X_i \in \real$, $i = 1, \ldots, n$, according to the joint density function $f_{X_1,\ldots, X_n}$. $\cbrace{X_i}_i$ are arbitrarily correlated. The resulting random vector is denoted by $\vec{X} \in\real^n$. Observe that $n$ may not be equal to $N$. This setup encompasses the case in which some of the sources are co-located, or some physical sources generate random variables of higher dimension, by grouping some of the sources into a source node in $\mathcal{S}$. Thus, when appropriate we may refer to a source node $s \in \mathcal{S}$ to represent the collection of virtual sources which are co-located at $s$ (c.f. Experiments section~\ref{sec:experiment}).

We model each link in the network  as a set of parallel channels. More precisely, two nodes $v_i$ and $v_j$ in the network are connected via $k_{i,j}$ parallel noisy channels. On each link $\paren{i,j}$ in the network, $v_i$ may transmit any signal $\vec{W}_{i,j}$ in $\real^{k_{i,j}}$, subject to the average power constraint on the link, i.e. $\mathbb{E}[\vec{W}_{i,j}^2] \leq p_{i,j}$. The signal $\vec{W}_{i,j}$ on each link then gets corrupted by noise. Node $v_j$ receives a corrupted signal $\vec{Y}_{i,j} \in \real^{k_{i,j}}$. In the special case of independent zero-mean Additive White Gaussian Noise (AWGN) channels, the node $v_j$ receives $\vec{Y}_{i,j} = \vec{W}_{i,j} + \vec{N}_{i,j}$, where each element of the $k_{i,j}$-dimensional vector $\vec{N}_{i,j}$ is a zero-mean Gaussian random variable with variance $\sigma_{i,j}^2$. Note that this setup models wireless point-to-point communication where the $k_{i,j}$ independent channels are obtained by constructing orthogonal sub-channels from the available bandwidth \cite{cover2012elements}.

We study the multi-cast problem where information generated at the source nodes must be communicated to the destination nodes. At each destination node $t \in \mathcal{D}$, an estimate $\vec{X}^t$ of the source $\vec{X}$ is reconstructed. Performance of the multi-casting scheme can be evaluated by a tuple of distortion measure $\delta_t$s, with each one of which defined between the source $\vec{X}$ and the estimation $\vec{X}^t$ at a destination $t$. 

\begin{figure}
    \centering
    \includegraphics[width=1 \columnwidth]{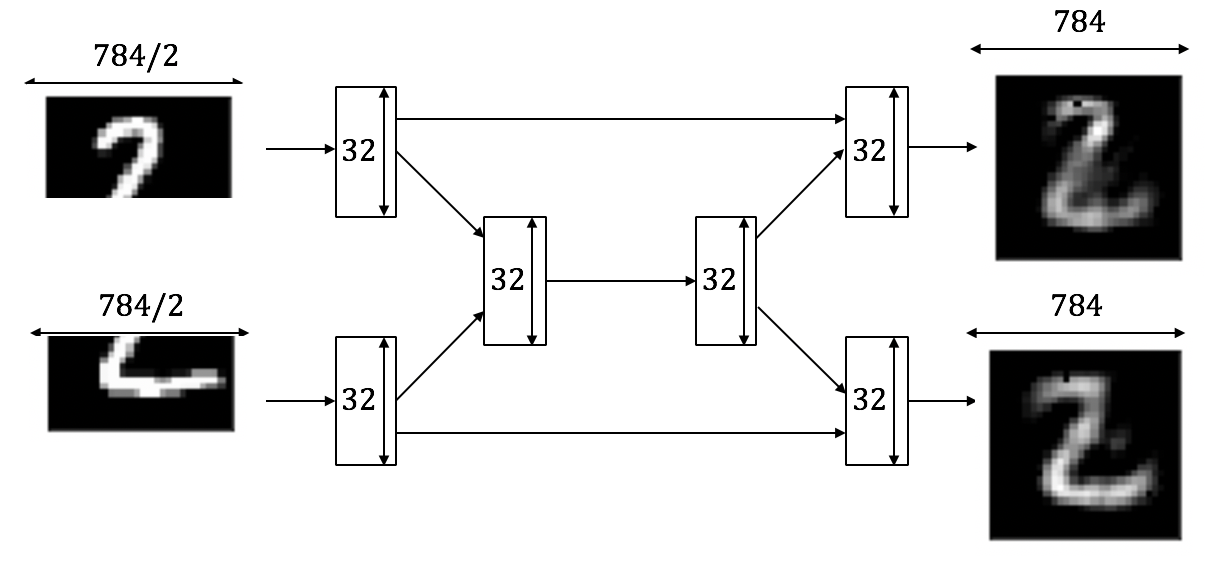}
    \caption{An end-to-end NN example. There are $n=784$ virtual sources located at $N=2$ source nodes, and there are $M=2$ destination nodes. Each link consists of 32 channels. Each node has an inner NN, to construct and decode network codes. For example, the inner NN at each destination node has input dimension $64$ and output dimension $784$. Each link is implemented with a non-trainable NN layer of width $32$.\squeezeup{15}}
    \label{fig:healthy_butterfly}
\end{figure}

\section{Neural Network Coding}\label{sec:method}

In NNC, we design the channel inputs at the source nodes, and at the intermediate nodes jointly -- this makes NNC a joint source and network coding scheme.
Existing joint source and network coding schemes, e.g., \cite{ho2006random,wu2009practical,lee2007minimum,koetter2003algebraic}, 
assume error-free point-to-point transmission on each link, and focus on the network layer operations. 
The physical layer then relies on a separate channel coding scheme with potentially high latency, as it is assumed that each link is employing an error correcting code with a large block length.
In contrast, in NNC the signal inputs are directly transmitted over the noisy channels, i.e. there are no underlying physical layer codes.
As such, the communication problem described in Section~\ref{sec:model} can be decomposed into three phases as shown in Figure~\ref{fig:general}(a): the encoding phase, the transmission phase and the decoding phase. 
NNC operates in a \textit{one-shot} manner over all three phases. 
In the encoding phase, real-valued signals at the source nodes are directly mapped into network codes. 
The length of a network code $\vec{W}_{i,j}$ is designed to match the number of independent channels $k_{i,j}$ consisted in link $(i,j)$. 
$\vec{W}_{i,j}$ can therefore be transmitted concurrently through the noisy channels over link $(i,j)$.
In the transmission phase, network codes $\vec{W}_{i,j}$s are directly constructed at node $v_i$ from the incoming noise-corrupted network codes $\cbrace{\vec{Y}_{l,i}: l \in c(i)}$, where $c(i)$ is the set of direct predecessors of $v_i$. 
In the decoding phase, each destination node reconstructs the transmitted signals directly from the noise-corrupted network codes it receives. NNC does not involve any long block-length source or channel coding techniques, and therefore is free of their associated latency penalty and computational cost.

Note that by picking a non-linear activation, the resulting joint source and network code is non-linear by design. As mentioned in~Section~\ref{sec:intro}, the non-linearity in codes may be crucial in constructing efficient codes for the problem at hand.
We design the network code from node $v_i$ to node $v_j$ by constructing a NN with input dimension $d^i_{in}$ and output dimension $k_{i,j}$. When $v_i \centernot\in \mathcal{S}$, $d^i_{in} = \sum_{l \in c(i)} k_{l,i}$. When $v_i \in \mathcal{S}$, $d^i_{in}$ is the dimension of signal generated at $v_i$. During a transmission, the concatenation of noise-distorted network codes received at $v_i$, $\{\vec{Y}_{l,i}: l \in c(i) \}$, is fed into the NN if $v_i \centernot\in\mathcal{S}$. Or the generated signal is fed into the NN if $v_i\in\mathcal{S}$. The NN output is the network code $\vec{W}_{i,j}$ to be transmitted over link $(i,j)$. 

Similarly, we reconstruct the input signal as $\vec{X}^t$ by decoding the received noise-distorted network codes with a NN at each destination node $t$. 
Note that NNs at destination nodes are low-complexity decoders, since each layer of a NN is an affine transformation followed by an element-wise non-linear function.
We say that the set of functions for constructing and decoding network codes at each node specifies a \textit{NNC policy} for the communication system, if each of them can be represented as a NN. 
Under a NNC policy, the end-to-end resulting encoding-decoding can be seen as a series of NNs connected via noisy links, as given by the physical network topology.
It will be convenient to represent those the noisy links by NN-layers as well, with the main difference that those layers have fixed (non-trainable) parameters which will correspond to the channel statistics.
Thus, under a NNC policy, we construct an end-to-end NN, where some of the layers have non-trainable parameters. 
The end-to-end NN has physical topology of the communication system embedded, and has NNs which are used for constructing and decoding network code as its sub-graphs.
We refer to the NNs for constructing and decoding network code as \textit{inner NNs} henceforth. 
Overall, there are $N$ input layers and $M$ output layers in the end-to-end NN.
Each input layer has width equal to the dimension of source generated at the node. 
All output layers have width $n$.  
An illustration of an end-to-end NN is given in Figure~\ref{fig:healthy_butterfly}. 

With $\vX$ partitioned and fed into the input layers, the outputs of the end-to-end NN simulate $\{ \vec{X}^t \}_t$, the reconstruction at destination nodes under current NNC policy. 
Recall that $\delta_t$ is the distortion measure between the source $\vec{X}$ and the estimation $\vec{X}^t$ at a destination node $t$, as defined in Section~\ref{sec:model}.
Parameters $\{ \theta_l \}_l$ of the NNC policy are initialized randomly and are trained to minimize 
\begin{align}\label{eq:lag_obj}
    \sum_t \delta_t + \sum_{i,j}\lambda_{i,j}\mathbb{E}[\vec{W}_{i,j}^2].
\end{align}
over a large data set sampled from $f_{X_1,\ldots, X_n}$. Note the optimization problem is the Lagrangian relaxation of the problem discussed in Section~\ref{sec:model}. 
We control the transmission power implicitly by penalizing power in the objective function through $\cbrace{\lambda_{i,j}}$: The larger $\lambda_{i,j}$ is, the more the power on link $(i,j)$ is penalized.
The parameters of the NN policy can be trained and tested offline\footnote{For the best performance, efforts are in general required to optimize over the choice of hyper-parameters as well, as is the case in other applications of NNs. Hyper-parameters, such as the number of layers and the activation functions in every inner NN, can also be tuned and tested offline before implementation.}, using a variety of available tools, e.g., \cite{chollet2015keras}, \cite{tensorflow2015-whitepaper}. 
Note that for the simple topology of a single hop link, NNC reduces to deep JSCC in \cite{bourtsoulatze2019deep} with soft control on transmission power.





\begin{figure*}[t!]
\centering
\includegraphics[width=2 \columnwidth]{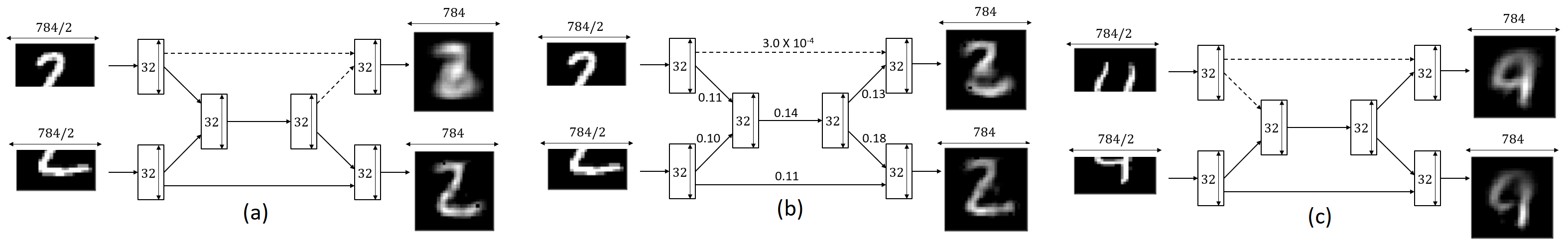}
\caption{Illustration of network codes construction in NNC. (a) Butterfly network where the top destination node has a weak receiver. Both incoming links to the top receiver suffer from low SNR. (b) Butterfly network with a weak link from the top source node to the top destination node. Transmission on that link suffers from low SNR. The numbers on the links represent the average power per transmission on each link. (c) Butterfly network where the top source node has a weak transmitter. Both outgoing links from the top source node suffer from low SNR.}%
\label{fig:injured}
\end{figure*}

\section{Performance Evaluation}\label{sec:experiment}

We studied the performance of NNC by experimenting with multi-casting an MNIST image \cite{lecun1998mnist} over a butterfly network, as shown in Figures~\ref{fig:healthy_butterfly},\ref{fig:injured}. 
In this setup, there are two source nodes ($N = 2$) and two destination nodes ($M = 2$). 
A normalized MNIST image, with pixel values between $0$ and $1$, is split between the two source nodes, such that each source node observes only one half of the image.
In other words, $392$ out of $n = 28 \times 28$ virtual sources (pixels) are co-located at each source node, where the top $392$ pixels are located at the first source node, and the rest at the second. 
Each link in the butterfly network consists of $32$ independent parallel AWGN channels ($k_{i,j}=32 ,\ \forall i,j$), with zero-mean noise of variance $10^{-4}$. 
We experimented over different power constraints, resulting in different SNR over the links. 
Performance is evaluated by the peak signal to noise ratio (pSNR) at each destination node, defined as
\begin{align}
    pSNR_t = 10 \log_{10} (\frac{\max\{X_i\}^2}{\mathbb{E}[(\vec{X}^t - \vec{X})^2]}),
\end{align}
where $\max\{X_i\}$ is the largest value of input pixels. 
The choice of pSNR as a distortion measure is natural for images \cite{welstead1999fractal}. 
The pSNR is essentially a normalized version of the MSE and can be used for performance comparison between inputs with different pixel ranges.

In each experiment, we learnt a NNC policy with every inner NN set to be two-layer fully-connected with activation function ReLU: $f(x)=x^{+}=\max(0,x)$. 
Note that the hyper-parameters here may not be optimal, but the results still serve as a proof of concept. 
A NNC policy is trained to minimize
\begin{align}\label{objective}
    \min_{\{\theta_l\}_l} \sum_{t=1}^{2}H_b(\vec{X},\vec{X}^t) + \sum_i\lambda_{i,j}\mathbb{E}[\vec{W}_{i,j}^2],
\end{align}
where $H_b(\vec{X},\vec{X}^t)$ is the binary cross entropy between the original image and the reconstructed image at destination node $t$. Note the dependence of $H_b$ and $\vec{W}_{i,j}$ on $ \{ \theta_l \}_l$ is omitted in the expression for simplicity. We use binary cross entropy $H_b(\vec{X},\vec{X}^t)$, defined as
\begin{align*}
    H_b(\vec{X},\vec{X}^t) = -\frac{1}{n}\sum_i X_i \log(X^t_i) +  (1 - X_i) \log(1 - X^t_i), 
\end{align*} 
in the objective function instead of pSNR as an engineering tweak to speed up the training process. 
Each NNC policy is learnt through training over 60000 MNIST training images for 150 epochs, and is tested on 10000 MNIST testing images. 
Note that the training set and test set are disjoint. 
We implemented
the NN architecture in  Keras\cite{chollet2015keras} with TensorFlow\cite{tensorflow2015-whitepaper} backend. 
Adadelta optimization framework \cite{zeiler2012adadelta} is used with learning rate of 1.0 and a mini-batch size of 32 samples.

In our experiments, we studied power-distortion trade-off of NNC under a variety of network conditions. Different network condition is enforced by different choice of $\cbrace{\lambda_{i,j}}$: The higher $\lambda_{i,j}$ is, the less power is expected to be sent on link $\paren{i,j}$. We call a link in the network ``weak" if transmission on the link suffers from lower SNR compared to other links. Weak links are denoted by dashed arrows in the diagrams. We say a node has weak transmitter/receiver if all its outgoing/incoming links are weak. The nodes and links in the network are ``equally strong" if power sent on all links are penalized with the same $\lambda$. We first qualitatively studied NNC's performance in heterogeneous networks. We then quantitatively studied NNC's performance in homogeneous networks. We analyzed its power allocation strategy, and demonstrated the benefit of allowing non-linearity in network code construction and having a joint coding scheme through comparison.

\subsection{Heterogeneous Networks}

Our first set of experiments studies the performance of NNC under a variety of network conditions. For each network condition, we visualize in Figures~\ref{fig:healthy_butterfly},\ref{fig:injured} the power-distortion trade-off by showing an instance of test image reconstruction. 

Figure~\ref{fig:healthy_butterfly} shows the transmission of an MNIST image over a butterfly network where all nodes and links are equally strong. Power sent on every link ($i,j$) is equally penalized with a small penalization weight $\lambda_{i,j}$ in this experiment, resulting in high SNR on all links, so image reconstruction at both destination nodes is successful.

Figure~\ref{fig:injured}(a) shows one network where the top destination node has a weak receiver. Incoming signals to the top receiver suffer from a low SNR. As a result, the top destination node reconstructs the image poorly, while the bottom destination node performs a much better reconstruction. 

Figure~\ref{fig:injured}(b) shows another type of such network where only the top link is weak. As a result, the reconstruction of the upper half of the image at the top destination node is poorer than that at the bottom destination node. Both destination nodes reconstruct the lower half of the image well, as both links from the bottom source node have high SNR. Note that the middle link carries enough information about both halves of the image to allow the top destination node to partially reconstruct the top half of the image. This is a result of the network adjusting to the top link suffering from low SNR, and therefore uses its power in other links.

The last example of a heterogeneous link conditions is depicted in Figure~\ref{fig:injured}(c). In this network, the top source node has a weak transmitter and both outgoing links from the top source node suffer from low SNR. As a result, both destination nodes cannot reconstruct the upper half of the image well. Note that both destination nodes reconstruct well the lower half of the image, as SNR on both outgoing links from the bottom source are high. Furthermore, we notice that the bottom destination node performs better on the top half of the image than the top destination node. This might be explained by the bottom destination node being able to infer the top half of the image better than the top destination node with knowledge from the lower half of the image, as the two halves of the image are correlated.

\subsection{Homogeneous Networks}

\begin{figure}
    \centering
    \includegraphics[width=.95\columnwidth]{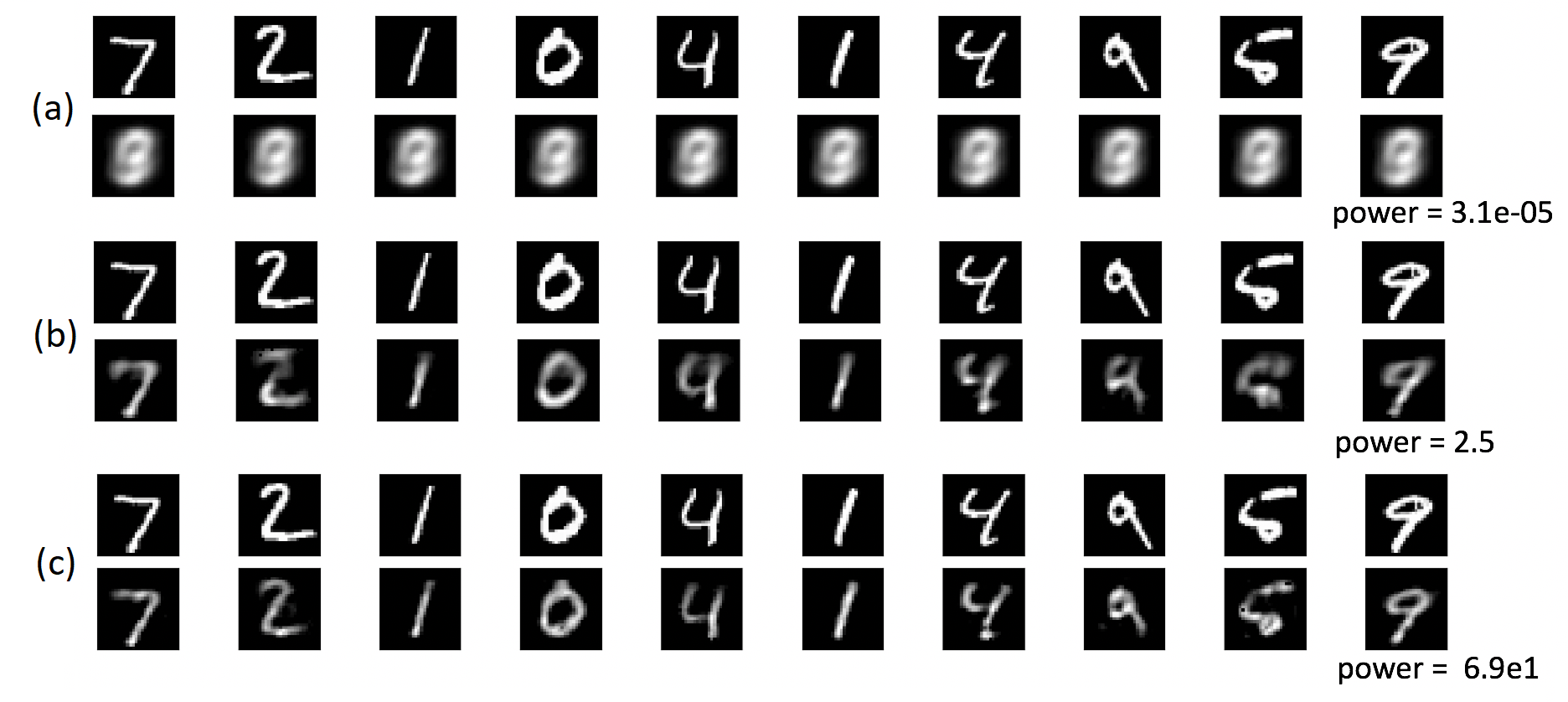}
    \caption{Illustration of transmission quality with different power. The first row of each subfigure is the original image, and the following row is reconstruction at one of the two destination nodes of a butterfly network. The two reconstructed images at both destination nodes were similar. (a) low power, (b) medium power, (c) high power.\squeezeup{5}}
    \label{fig:illustration_quality}
\end{figure}

Our second set of experiments studies the power-distortion trade-off of NNC on a homogeneous butterfly network, where all nodes and links are equally strong. 
Note that since the noise on each link is fixed, this is equivalent to studying the SNR-distortion trade-off.
The transmission power per image is implicitly controlled by the value of $\lambda$. 
As expected, the quality of the reconstruction improves as transmission power increases, from Figure~\ref{fig:illustration_quality}(a) (low power) to Figure~\ref{fig:illustration_quality}(b) (medium power), and finally to Figure~\ref{fig:illustration_quality}(c) (high power). 
Note that when the transmission power is almost forced to be zero, as shown in Figure~\ref{fig:illustration_quality}(a), both destination nodes reconstruct the average of training data. 
It is essentially the best possible reconstruction as no information flows from the sources to destinations.
In addition, Table~\ref{tab:waterfilling} illustrates the power allocation of NNC with different power budget. When limited on power, i.e., $\lambda$ is large, NNC prefers to send information on less channels with higher SNR rather than spreading energy over more channels. 
Indeed, by not allocating power to some channels, the power budget can be used to improve the quality of the channels which carry information.
This is in line with the intuition from the water-filling algorithm\cite{proakis1994communication}.

\begin{table}
\centering
    \vspace{1em}
        \begin{tabular}{l l l l}
        \hline
        {$\lambda$} & {P} & {nonzero} & {$>$ 3dB} \\ \hline
        $10^{-5}$ & 31.4 & $76\%$ & $59\%$\\
        $10^{-4}$ & 6.2 & $67\%$ & $40\%$\\
        $10^{-3}$ & 1.1 & $28\%$ & $8\%$\\
        $10^{-2}$ & 0 & $0\%$ & $0\%$\\
        \end{tabular}
        \caption{Illustration of power allocation in a homogenous butterfly network. \textit{$\lambda$} is the power penalization parameter.  \textit{$P$} is the average transmission power per image. 
        \textit{nonzero} is the average percentage of all channels with nonzero power allocation per image. 
        \textit{$>3dB$} is the average percentage of channels with SNR greater than $3dB$.}
        \label{tab:waterfilling}
\end{table}

We compared the performance of NNC with two baseline methods. 
The first competitor is a \textit{linear} analog of NNC, the Analog Network Coding scheme (ANC)\cite{maric2010analog}: each node amplifies and forwards the sum of its inputs.
All amplification factors are the same and the destination nodes decode knowing the amplification factor and the network topology. 
Note that a $28\times28$ MNIST image can be sent by NNC in one-shot, but has to be sent over the network in $13$ transmissions by ANC, as there is no compression scheme in the ANC baseline and thus at most $64$ pixels can be sent in a single transmission. 
All distortion in the reconstruction comes from the noise in the channel under the ANC baseline.

\begin{figure}
    \centering
    \includegraphics[width=1 \columnwidth]{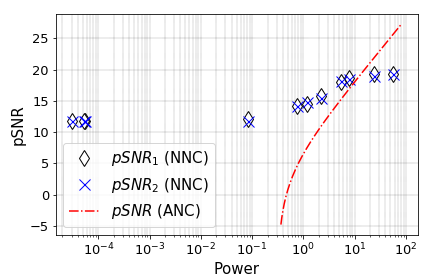}
    \caption{pSNR at the two destination nodes, using NNC and ANC, as a function of the average transmission power per image. The pSNR at two destination nodes are equal under ANC. Note that NNC performs better in lower transmission power.\squeezeup{10}}
    \label{fig:AF_competitor}
\end{figure}

Figure~\ref{fig:AF_competitor} compares the performance of NNC and the ANC baseline. Transmission power in ANC is controlled by amplification factor. 
The average transmission power and pSNR at both destination nodes in Figure~\ref{fig:AF_competitor} are averaged over 300 runs over the test set. 
Note that performances of both scheme are ``symmetric" at destination nodes, reconstructing images with similar pSNR. 
Such ``symmetric" performance can be expected since the network is homogeneous. 
Overall, NNC outperforms the ANC baseline when transmission power is low, and the ANC baseline outperforms NNC when transmission power is high.
This is consistent with \cite{maric2010analog} which shows that ANC is in fact capacity-achieving in the high-SNR regime.


The second competitor, the JPEG baseline, is a scheme that separates source coding from network coding: images are compressed through the JPEG compression algorithm at source nodes, and are then transmitted by capacity-achieving network codes through error-free channel codes over the network. 
Distortion in the reconstruction under the JPEG baseline only comes from compression in source coding, as the transmission is assumed to be error-free. 
Notice that the JPEG baseline has potentially a high latency due to error-free channel coding, which operates on a large block length. 
 
In our experiments, the JPEG baseline reconstructs high quality images with impractically high power. 
With the JPEG baseline, average pSNR between a reconstructed image and the original image ranges from 17 to 46.
However, the minimum power threshold for using the JPEG competitor is tremendously high as $10^{51}$ per image. 
The need of such high transmission power is explained by JPEG algorithm hardly compressing MNIST images. 
Before compression, each half MNIST image is 392 bytes. After compression, the average file size of half images ranges from 350 bytes to 580 bytes for different qualities. 
For small images like MNIST, the overhead of the JPEG algorithm is significant. 
The same problem exists for other compression schemes like JPEG2000.
Insufficient compression by JPEG is a representative example of how traditional schemes may lack the ability to adapt their rates to different communication scenarios.

\section{Conclusion}\label{sec:conclusion}

In this paper, we proposed a novel way of constructing network codes using NN. Our scheme, NNC, provides a practical solution for multi-casting arbitrarily correlated signals over networks of arbitrary typologies in an application-specific manner. NNC can be easily learnt and tested offline and implemented in a distributed fashion. We examined the performance of NNC under a variety of network conditions through experiments. 

Three possible extensions of the problem arise naturally. First of all, a separate NNC policy can be learnt for each task when multiple tasks exist asynchronously in a system. Thanks to the simplicity of NNC implementation, multiple functions can be implemented at one node. Signals can be designed with a flag piggybacked to trigger task-specific functions for constructing and decoding network codes at each node. Second, NNC can also be extended with little effort to the theoretically hard case when each destination node has to reconstruct a subset of the sources \cite{yan2007capacity,song2003zero}; in the multi-casting problem all source signals must be reconstructed at every destination node. 
Third, a functional variation of NNC can be applied when the destination nodes are interested in functions of the input, rather than the input itself. 
For example, in aforementioned experiments, destination nodes can be interested in classifying which hand-written digit is sent, rather than reconstructing the image itself. These extensions are not specifically discussed in this paper due to space constraints.

Future work includes testing NNC's performance with a variety of source types, network topologies and channel models. For example, an erasure channel can be readily implemented by a drop-out NN layer. Previous studies on point-to-point transmission, such as text transmission in \cite{farsad2018deep} and transmission over a multi-path channel in \cite{felix2018ofdm}, can be extended with NNC scheme to transmission over a network. An additional direction could be extending NNC over elements drawn from a finite field (as opposed to real numbers), which would allow NNC to be used in a digital domain. Previous works on the quantization of NNs, e.g., \cite{jacob2018quantization}, can be referred to for this extension. 

\section*{Acknowledgment}
The authors would like to thank Yushan Su and Alejandro Cohen for their technical help and constructive comments. 
\appendices

\begin{spacing}{0.98}
\bibliographystyle{IEEEtran}
\bibliography{paper}

\begin{thebibliography}{10}
\providecommand{\url}[1]{#1}
\csname url@samestyle\endcsname
\providecommand{\newblock}{\relax}
\providecommand{\bibinfo}[2]{#2}
\providecommand{\BIBentrySTDinterwordspacing}{\spaceskip=0pt\relax}
\providecommand{\BIBentryALTinterwordstretchfactor}{4}
\providecommand{\BIBentryALTinterwordspacing}{\spaceskip=\fontdimen2\font plus
\BIBentryALTinterwordstretchfactor\fontdimen3\font minus
  \fontdimen4\font\relax}
\providecommand{\BIBforeignlanguage}[2]{{%
\expandafter\ifx\csname l@#1\endcsname\relax
\typeout{** WARNING: IEEEtran.bst: No hyphenation pattern has been}%
\typeout{** loaded for the language `#1'. Using the pattern for}%
\typeout{** the default language instead.}%
\else
\language=\csname l@#1\endcsname
\fi
#2}}
\providecommand{\BIBdecl}{\relax}
\BIBdecl

\bibitem{o2016learning}
T.~J. O'Shea, K.~Karra, and T.~C. Clancy, ``Learning to communicate: Channel
  auto-encoders, domain specific regularizers, and attention,'' in \emph{Proc.
  of IEEE Int. Symp. on Signal Processing and Information Technology (ISSPIT)},
  Dec. 2016, pp. 223--228.

\bibitem{o2017introduction}
T.~J. O'Shea and J.~Hoydis, ``An introduction to deep learning for the physical
  layer,'' \emph{IEEE Transactions on Cognitive Communications and Networking},
  vol.~3, no.~4, pp. 563--575, 2017.

\bibitem{felix2018ofdm}
A.~Felix, S.~Cammerer, S.~D{\"o}rner, J.~Hoydis, and S.~Ten~Brink,
  ``{OFDM}-autoencoder for end-to-end learning of communications systems,'' in
  \emph{Proc. IEEE Int. Workshop Signal Proc. Adv. Wireless Commun. (SPAWC)},
  2018.

\bibitem{bourtsoulatze2019deep}
E.~Bourtsoulatze, D.~B. Kurka, and D.~G{\"u}nd{\"u}z, ``Deep joint
  source-channel coding for wireless image transmission,'' \emph{IEEE
  Transactions on Cognitive Communications and Networking}, 2019.

\bibitem{farsad2018deep}
N.~Farsad, M.~Rao, and A.~Goldsmith, ``Deep learning for joint source-channel
  coding of text,'' in \emph{Proc. IEEE Int. Conf. on Acoustics, Speech and
  Signal Processing(ICASSP)}.\hskip 1em plus 0.5em minus 0.4em\relax IEEE,
  2018, pp. 2326--2330.

\bibitem{slepian1973noiseless}
D.~Slepian and J.~Wolf, ``Noiseless coding of correlated information sources,''
  \emph{IEEE Trans. Inf. theory}, vol.~19, no.~4, pp. 471--480, 1973.

\bibitem{csiszar1982linear}
I.~Csiszar, ``Linear codes for sources and source networks: Error exponents,
  universal coding,'' \emph{IEEE Trans. Inf. theory}, vol.~28, no.~4, pp.
  585--592, 1982.

\bibitem{song2001network}
L.~Song and R.~W. Yeung, ``Network information flow-multiple sources,'' in
  \emph{Proceedings. 2001 IEEE Int. Sym. Inf. Theory}, 2001, p. 102.

\bibitem{ho2006random}
T.~Ho, M.~M{\'e}dard, R.~Koetter, D.~R. Karger, M.~Effros, J.~Shi, and
  B.~Leong, ``A random linear network coding approach to multicast,''
  \emph{IEEE Trans. Inf. theory}, vol.~52, no.~10, pp. 4413--4430, 2006.

\bibitem{maierbacher2009practical}
G.~Maierbacher, J.~Barros, and M.~M{\'e}dard, ``Practical source-network
  decoding,'' in \emph{2009 6th International Symposium on Wireless
  Communication Systems}.\hskip 1em plus 0.5em minus 0.4em\relax IEEE, 2009,
  pp. 283--287.

\bibitem{coleman2005towards}
T.~P. Coleman, M.~M{\'e}dard, and M.~Effros, ``Towards practical
  minimum-entropy universal decoding,'' in \emph{Data Compression
  Conference}.\hskip 1em plus 0.5em minus 0.4em\relax IEEE, 2005, pp. 33--42.

\bibitem{ramamoorthy2006separating}
A.~Ramamoorthy, K.~Jain, P.~A. Chou, and M.~Effros, ``Separating distributed
  source coding from network coding,'' \emph{IEEE/ACM Trans. Netw.}, vol.~14,
  no.~SI, pp. 2785--2795, 2006.

\bibitem{wu2009practical}
Y.~Wu, V.~Stankovic, Z.~Xiong, and S.-Y. Kung, ``On practical design for joint
  distributed source and network coding,'' \emph{IEEE Trans. Inf. theory},
  vol.~55, no.~4, pp. 1709--1720, 2009.

\bibitem{lee2007minimum}
A.~Lee, M.~M{\'e}dard, K.~Z. Haigh, S.~Gowan, and P.~Rubel, ``Minimum-cost
  subgraphs for joint distributed source and network coding,'' in \emph{Proc.
  NETCOD}, 2007.

\bibitem{goodfellow2016deep}
I.~Goodfellow, Y.~Bengio, and A.~Courville, \emph{Deep learning}.\hskip 1em
  plus 0.5em minus 0.4em\relax MIT press, 2016.

\bibitem{wallace1992jpeg}
G.~K. Wallace, ``The {JPEG} still picture compression standard,'' \emph{IEEE
  Transactions on Consumer Electronics}, vol.~38, no.~1, pp. xviii--xxxiv,
  1992.

\bibitem{cormen2009introduction}
T.~H. Cormen, C.~E. Leiserson, R.~L. Rivest, and C.~Stein, \emph{Introduction
  to algorithms}.\hskip 1em plus 0.5em minus 0.4em\relax MIT press, 2009.

\bibitem{cover2012elements}
T.~M. Cover and J.~A. Thomas, \emph{Elements of information theory}.\hskip 1em
  plus 0.5em minus 0.4em\relax John Wiley \& Sons, 2012.

\bibitem{koetter2003algebraic}
R.~Koetter and M.~M{\'e}dard, ``An algebraic approach to network coding,''
  \emph{IEEE/ACM Transactions on Networking (TON)}, vol.~11, no.~5, pp.
  782--795, 2003.

\bibitem{chollet2015keras}
F.~Chollet \emph{et~al.}, ``Keras,'' \url{https://keras.io}, 2015.

\bibitem{tensorflow2015-whitepaper}
\BIBentryALTinterwordspacing
M.~Abadi \emph{et~al.}, ``{TensorFlow}: Large-scale machine learning on
  heterogeneous systems,'' 2015, software available from tensorflow.org.
  [Online]. Available: \url{http://tensorflow.org/}
\BIBentrySTDinterwordspacing

\bibitem{lecun1998mnist}
Y.~LeCun, ``The mnist database of handwritten digits,'' \emph{http://yann.
  lecun. com/exdb/mnist/}, 1998.

\bibitem{welstead1999fractal}
S.~T. Welstead, \emph{Fractal and wavelet image compression techniques}.\hskip
  1em plus 0.5em minus 0.4em\relax SPIE Optical Engineering Press Bellingham,
  Washington, 1999.

\bibitem{zeiler2012adadelta}
M.~D. Zeiler, ``Adadelta: an adaptive learning rate method,'' \emph{arXiv
  preprint arXiv:1212.5701}, 2012.

\bibitem{proakis1994communication}
J.~G. Proakis, M.~Salehi, N.~Zhou, and X.~Li, \emph{Communication systems
  engineering}.\hskip 1em plus 0.5em minus 0.4em\relax Prentice Hall New
  Jersey, 1994, vol.~2.

\bibitem{maric2010analog}
I.~Maric, A.~Goldsmith, and M.~M{\'e}dard, ``Analog network coding in the
  high-{SNR} regime,'' in \emph{2010 Third IEEE International Workshop on
  Wireless Network Coding}.\hskip 1em plus 0.5em minus 0.4em\relax IEEE, 2010,
  pp. 1--6.

\bibitem{yan2007capacity}
X.~Yan, R.~W. Yeung, and Z.~Zhang, ``The capacity region for multi-source
  multi-sink network coding,'' in \emph{2007 IEEE Int. Sym. Inf. Theory}.\hskip
  1em plus 0.5em minus 0.4em\relax IEEE, 2007, pp. 116--120.

\bibitem{song2003zero}
L.~Song, R.~W. Yeung, and N.~Cai, ``Zero-error network coding for acyclic
  networks,'' \emph{IEEE Trans. Inf. theory}, vol.~49, no.~12, pp. 3129--3139,
  2003.

\bibitem{jacob2018quantization}
B.~Jacob, S.~Kligys, B.~Chen, M.~Zhu, M.~Tang, A.~Howard, H.~Adam, and
  D.~Kalenichenko, ``Quantization and training of neural networks for efficient
  integer-arithmetic-only inference,'' in \emph{Proc. IEEE Computer Vision and
  Pattern Recognition}, 2018, pp. 2704--2713.

\end{thebibliography}
\end{spacing}

\end{document}